\documentclass[aps,pra,superscriptaddress,amsfonts,longbibliography]{revtex4}

\usepackage{amsmath}
\usepackage{amssymb}
\usepackage{graphicx}
\usepackage{bm}
\usepackage{array}
\usepackage{dcolumn}
\usepackage{epstopdf}
\usepackage{xcolor}

\newcommand{\ket}[1]{{| {#1} \rangle}}

\begin{document}

\title{Ion-Based Characterization of Laser Beam Profiles for Quantum Information Processing}

\author{Ilyoung Jung, Frank G. Schroer and Philip Richerme}

\affiliation{Indiana University Department of Physics, Bloomington, Indiana 47405, USA}
\affiliation{Indiana University Quantum Science and Engineering Center, Bloomington, Indiana 47405, USA}

\date{\today}

\begin{abstract}
Laser-driven operations are a common approach for engineering one- and two-qubit gates in trapped-ion arrays. Measuring key parameters of these lasers, such as beam sizes, intensities, and polarizations, is central to predicting and optimizing gate speeds and stability. Unfortunately, it is challenging to accurately measure these properties at the ion location within an ultra-high vacuum chamber. Here, we demonstrate how the ions themselves may be used as sensors to directly characterize the laser beams needed for quantum gate operations. Making use of the four-photon Stark Shift effect in $^{171}$Yb$^+$ ions, we measure the profiles, alignments, and polarizations of the lasers driving counter-propagating Raman transitions. We then show that optimizing the parameters of each laser individually leads to higher-speed Raman-driven gates with smaller susceptibility to errors. Our approach demonstrates the capability of trapped ions to probe their local environments and to provide useful feedback for improving system performance.
\end{abstract}

\maketitle



\section{Introduction}
To process quantum information using trapped ions, it is necessary to address and manipulate the ion qubit states by coupling them to electromagnetic fields. In the majority of trapped ion experiments, the required fields are delivered via laser light directed towards the ion qubit array and tuned to a wavelength at or near the qubit resonance. This approach is already sufficient to generate single- and multi-qubit operations that form universal gate sets for quantum information processing \cite{nielsen2010quantum,divincenzo1995two,barenco1995elementary,cirac1995quantum,molmer1999multiparticle,schmidt2003realization,leibfried2003experimental}. Additionally, the ability for laser light to achieve fine frequency resolution compared to typical trapping frequencies and energy scales, as well as fine spatial resolution compared to inter-ion length scales, enables local single-qubit \cite{crain2014individual,lee2016engineering} and arbitrary two-qubit addressing \cite{wright2019benchmarking,pogorelov2021compact} that is central to quantum gate-model algorithms. For these reasons, laser-driven operations have been used in hundreds of quantum computing and simulation experiments over the past two decades \cite{wineland1998experimental,haffner2008quantum,bruzewicz2019trapped,monroe2021programmable}.

For many specific ion species, the qubit levels are encoded in Zeeman or hyperfine states with frequency splittings on the order of MHz to GHz \cite{wineland1998experimental,blinov2004quantum,olmschenk2007manipulation,benhelm2008experimental,keselman2011high,ruster2016long}. To address these states using laser light, and to generate the necessary momentum transfer for spin-motion coupling, two-photon stimulated Raman transitions have been employed for qubit manipulations \cite{ozeri2007errors}. In this scheme, a phase-coherent pair of Raman beams interacts with the trapped-ion array and drives transitions by coupling through a virtually excited state. Since the single- and two-qubit gate speeds depend on the alignments, spot sizes, and polarizations of these Raman beams, it is central to characterize and optimize these beam properties at the ion positions to predict and improve gate performance. Unfortunately for trapped-ion setups, traditional measurements of these quantities are made difficult since they must be made within an ultra-high vacuum chamber.

In this work, we demonstrate how trapped ions themselves may be used to probe and optimize the laser beams used to drive quantum gate operations. Our technique relies on measurement of the differential four-photon Stark shift, which results from driving the ion with mode-locked laser beams. Unlike conventional methods such as camera-based beam imaging or Rabi-frequency mapping, our technique leverages the high sensitivity of trapped ions to light-shift effects, enabling measurements directly at the ion within the trap. For each Raman beam independently, we observe a differential energy shift between qubit levels that scales quadratically with beam intensity and depends upon the beam polarization and magnetic field orientation. We use this signal to characterize the polarization, spot sizes, and alignments of each beam, which we then use to optimize the gate speed of two-photon Raman transitions. This approach, which is generalizable to any species of Zeeman or hyperfine qubit, turns the light-shift sensitivity of trapped ions into a tool for correcting potential sources of experimental imperfections.

\section{Methods}
\subsection{Experimental Setup}
\begin{figure}[h]
\begin{centering}
\includegraphics[width=6.5 cm,  trim=100 10 185 0, clip]{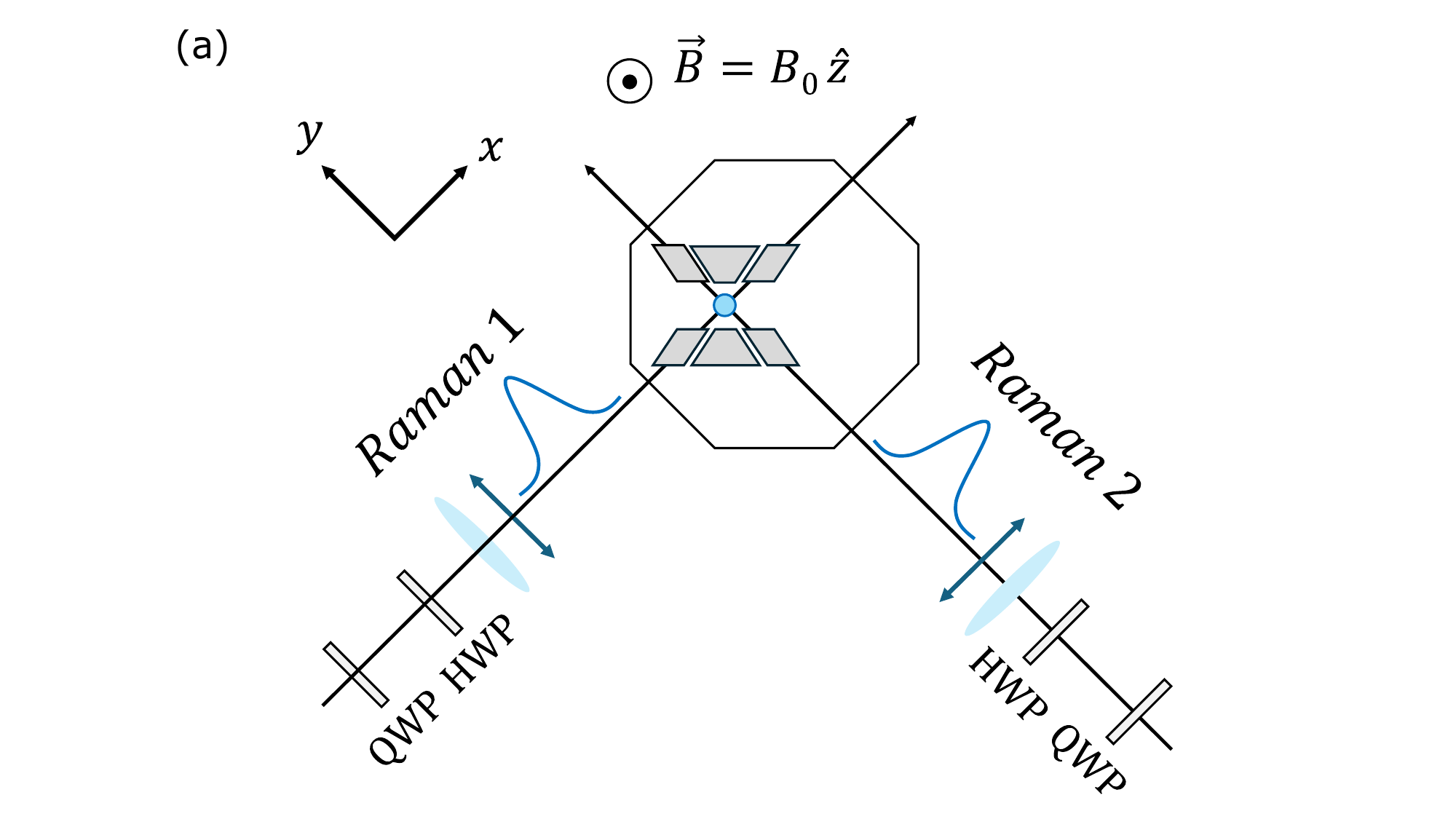}
\includegraphics[width=7.2 cm,  trim=100 0 130 10, clip]{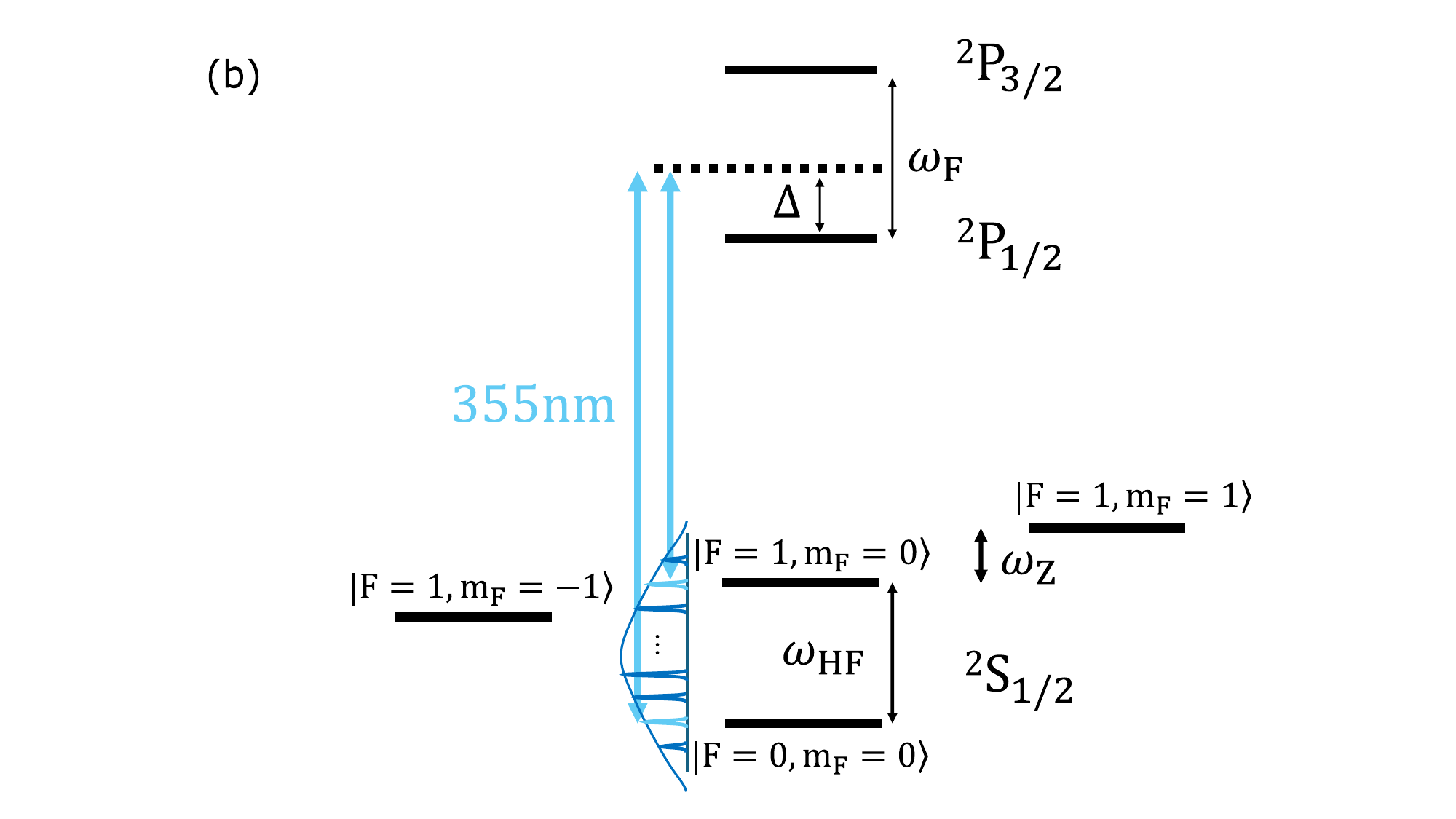}
\caption{Experimental Setup. (a) Diagram showing the ion trap within an ultra-high vacuum chamber. Also shown are the Raman beam configurations, ideal beam polarizations (double-headed arrows), and ideal magnetic field direction. QWP = Quarter-Wave Plate; HWP = Half-Wave Plate (b) Relevant energy level structure of $^{171}$Yb$^+$. Frequency comb components from a pair of 355 nm laser beams can drive Raman transitions, and each beam individually generates a differential four-photon Stark shift between qubit states.}
\vspace{-3pt}
\label{expsetup}
\end{centering}
\end{figure}

Experiments are performed using $^{171}$Yb$^+$ ions confined in an open-endcap linear rf trap \cite{xie2021open}. This choice of ion provides access to a hyperfine qubit, defined by the levels \mbox{$\ket{\downarrow}\equiv ^2$S$_{1/2}\ket{F=0,m_F=0}$} and \mbox{$\ket{\uparrow}\equiv ^2$S$_{1/2}\ket{F=1,m_F=0}$}, and split by a frequency $\omega_{HF}=2\pi\times 12.642812$~GHz. A magnetic field of 3.6 Gauss is oriented along the $\hat{z}$ direction (see Figure \ref{expsetup}(a)) to break the degeneracy of the $^2$S$_{1/2}~F=1$ manifold, resulting in a Zeeman shift of $\omega_Z = \pm2\pi\times 5$~MHz for the $^2$S$_{1/2}\ket{F=1,m_F=\pm 1}$ states (Figure \ref{expsetup}(b)). Ion cooling, initialization, and measurement are performed using resonant or near-resonant laser light at 369.5 nm \cite{olmschenk2007manipulation}.

We implement single and two-qubit gates by driving two-photon Raman transitions with a mode-locked laser at 355 nm (Figure 1). This laser, which outputs $\tau\approx15$ ps pulses at a repetition rate $\nu_{rep}\approx 80$~MHz, provides a frequency comb with $\sim$100 GHz bandwidth that spans the hyperfine qubit splitting. Resonant transitions between qubit levels, equivalent to single qubit rotations, are implemented by adjusting the relative frequency of each Raman beam such that the beatnote difference between comb teeth contains a component at $\omega_{HF}$. Similarly, two-qubit entangling operations are driven by generating a bichromatic beatnote between Raman beams that is near-resonant to the normal modes of the ion crystal \cite{molmer1999multiparticle,monroe2021programmable}.

Before interacting with the ions, each Raman beam is made to pass through a quarter-wave plate (QWP) and half-wave plate (HWP) so that its polarization may be fully controlled. In the ideal configuration (Figure \ref{expsetup}(a)), each Raman beam carries a linear polarization that is perpendicular to the external magnetic field as well as the polarization of the other beam. This `lin $\perp$ lin' configuration creates a polarization gradient at the ion \cite{metcalf1999laser}, which is used to drive the single- and two-qubit gates. We note that a single beam is unable to induce qubit transitions on its own, since the $\sigma^+$ and $\sigma^-$ contributions destructively interfere and $\pi$ transitions are forbidden by selection rules \cite{campbell2010ultrafast}.

\subsection{Four-photon Stark shift}
Although a single Raman beam cannot drive transitions between the $\ket{\downarrow}$ and $\ket{\uparrow}$ states, it may still cause a differential ac Stark shift in the qubit energy levels. For the standard (two-photon) case, the differential ac Stark shift is inversely proportional to the Raman beam detuning from each of the excited states. In $^{171}$Yb$^+$, there is a near-perfect cancellation of the two-photon Stark shift contributions from the $^2$P$_{1/2}$ and $^2$P$_{3/2}$ levels when the ion is irradiated with 355 nm light, since $\Delta \approx \omega_F/3$ (see Figure \ref{expsetup}(b)) \cite{wineland2003quantum,ozeri2007errors,tu2025precision}. For typical beam intensities, this results in a residual two-photon Stark shift of $\sim$10-100 Hz under the experimental conditions shown in Figure \ref{expsetup}(a) \cite{campbell2010ultrafast}.

When a mode-locked laser is used to drive Raman transitions, it induces a four-photon Stark shift that may be orders-of-magnitude larger than the usual two-photon effect \cite{lee2016engineering}. This is because a single Raman beam contains pairs of comb teeth separated approximately by $\omega_{HF}$, providing a near-resonant path for the four-photon virtual process. In \cite{lee2016engineering}, for instance, fourth-order Stark shifts of up to 10 MHz have been observed using 200 mW of 355 nm light focused to a 3~$\mu$m spot size, nearly 3 orders of magnitude larger than the corresponding two-photon Stark shift.

A full derivation of the four-photon Stark shift using perturbation theory is provided in \cite{lee2016engineering}; here we briefly summarize the relevant results. For a pair of comb teeth with frequencies $\omega_0$ and $\omega_1$ and polarizations $\hat{\epsilon}^0$ and $\hat{\epsilon}^1$, the Stark shift on state $\ket{n}$ is
\begin{equation}\label{eq:e4}
    E^{(4)}_{n} = \sum_{a\neq n}\frac{|\Omega_{n,a}|^2}{4\delta_{n,a}},
\end{equation}
where $\Omega_{n,a}$ is the two-photon Rabi frequency between states $\ket{n}$ and $\ket{a}$, and $\delta_{n,a}=(\omega_a-\omega_n)-(\omega_0-\omega_1)$ is the detuning of the comb beatnote from the relevant level splitting. For this analysis, we consider contributions from all states in the $^2$S$_{1/2}$ manifold (including the Zeeman states), leading to two-photon Rabi frequencies
\begin{align} \label{eq: Rabis}
\nonumber
    \Omega_{00,10} &= (\epsilon^{0}_{-}\epsilon^{1*}_{-}-\epsilon^{0}_{+}\epsilon^{1*}_{+})\Omega_{0}, \\
    \nonumber
    \Omega_{00,1-1} &= -(\epsilon^{0}_{-}\epsilon^{1*}_{\pi}+\epsilon^{0}_{\pi}\epsilon^{1*}_{+})\Omega_{0}, \\
    \nonumber
    \Omega_{00,11}&=(\epsilon^{0}_{+}\epsilon^{1*}_{\pi}+\epsilon^{0}_{\pi}\epsilon^{1*}_{-})\Omega_{0}, \\
    \nonumber
    \Omega_{10,1-1} &=(\epsilon^{0}_{-}\epsilon^{1*}_{\pi}+\epsilon^{0}_{\pi}\epsilon^{1*}_{+})\Omega_{0}, \\
    \Omega_{10,11} &= (\epsilon^{0}_{+}\epsilon^{1*}_{\pi}+\epsilon^{0}_{\pi}\epsilon^{1*}_{-})\Omega_{0},
\end{align} 
where transitions between states $\ket{n}$ and $\ket{a}$ are labeled by their values of $\ket{F,m_F}$. These Rabi frequencies depend upon the polarization components of each comb tooth, and the two-photon Rabi rate defined as:
\begin{equation}
    \Omega_{0} = \frac{g_{0}^{2}}{6}\left(\frac{1}{\Delta}+\frac{1}{\omega_{F}-\Delta}\right)
\end{equation}
with $g_{0}$ the single-photon resonant Rabi frequency between $S \rightarrow P$ \cite{campbell2010ultrafast,lee2016engineering}.
Generalizing Equation \ref{eq:e4} to include all pairs of comb teeth, and accounting for the envelope function of the frequency comb \cite{mizrahi2014quantum}, we write
\begin{equation}
    \label{eq:e4all}
    E^{(4)}_{n} = \sum_{a\neq n}  \frac{\Omega^{2}_{n,a}}{4}  \frac{\mathcal{C}_{n,a}}{\delta_{n,a}} ~~~~;~~~~
    \mathcal{C}_{n,a} = \sum_{k=-\infty}^{\infty}\frac{\text{sech}^2((j+k)\pi\nu_{rep}\tau)}{1-k(2\pi\nu_{rep})/\delta_{n,a}}
\end{equation}
with $j$ defined such that $|(\omega_a-\omega_n)-2\pi j \nu_{rep}|$ is minimized. Finally, the differential four-photon Stark shift is given by
\begin{equation}
\label{eq:dw4}
    \delta\omega^{(4)}=E_{10}^{(4)}-E_{00}^{(4)}
\end{equation}
where the $E_{n}^{(4)}$ shift for each level is calculated using Equation \ref{eq:e4all}.

In \cite{lee2016engineering}, the authors considered circular and linear Raman beam polarizations that maximized the four-photon Stark shift. Here, we extend the discussion to calculate the relevant Stark shifts for arbitrary polarizations and magnetic field orientations. To begin, we consider a vertically-oriented magnetic field and a Raman beam that carries pure linear polarization at angle $2\theta$ to the horizontal plane. With respect to the magnetic field, we may decompose the polarization vector as
\begin{equation}
    \label{eeq:polbasis}
    \hat{\epsilon} = \frac{\cos(2\theta)}{\sqrt{2}}\hat{\epsilon}_{-}+\text{sin}(2\theta) \hat{\epsilon}_\pi + \frac{\cos(2\theta)}{\sqrt{2}}\hat{\epsilon}_{+}
\end{equation}
For this polarization, we calculate the differential Stark shift using Eqs. \ref{eq: Rabis} and \ref{eq:e4all} as:
\begin{equation}
\label{eq:linpolss}
    \delta\omega^{(4)}(\theta) = \frac{\Omega^{2}_{0}}{8}\text{sin}^2(4\theta)\left ( \frac{C_{00,1-1}}{\delta_{00,1-1}} + \frac{C_{00,11}}{\delta_{00,11}} \right)
\end{equation}
From Equation \ref{eq:linpolss} we observe that the lin $\perp$ lin Raman beam polarizations ($\theta=0$), which is the `ideal' experimental configuration shown in Figure \ref{expsetup}(a), contributes zero differential Stark shift. 

We also calculate the four-photon Stark shift in the most general case, allowing for arbitrary magnetic field directions and arbitrary components of linear and circular polarization. This is motivated by the need to characterize real-world experimental implementations of Raman beam geometries, which may deviate from the ideal configuration shown in Figure \ref{expsetup}(a). We define an arbitrary external magnetic field, parameterized by polar and azimuthal angles $\alpha$ and $\beta$, with respect to the coordinate axes as
\begin{equation}
\vec{B}=B_0\left[\sin\alpha\cos\beta~\hat{x}+\sin\alpha\sin\beta~\hat{y}+\cos\alpha~\hat{z}\right]
\end{equation}
We also consider an input Raman beam that starts with horizontal polarization and is made to pass through a QWP at angle $\phi$ and a HWP at angle $\theta$. For this configuration, we derive the Raman beam polarization components as (see Appendix \ref{appxa}):
\begin{align}
    \label{eq:genepsilon}
    \nonumber
        \epsilon_- &= -\frac{1}{2}\left[(\cos\alpha\sin\beta-i\cos\beta)(\cos(2\phi-2\theta)+i\cos2\theta)+\sin\alpha(\sin(2\phi-2\theta)-i\sin2\theta)\right] \\[1em]
    \nonumber
    \epsilon_\pi &= -\frac{1}{\sqrt{2}}\left[\sin\alpha\sin\beta(\cos(2\phi-2\theta)+i\cos2\theta)-\cos\alpha(\sin(2\phi-2\theta)-i\sin2\theta)\right]\\[1em]
    \epsilon_+ &= -\frac{1}{2}\left[(\cos\alpha\sin\beta+i\cos\beta)(\cos(2\phi-2\theta)+i\cos2\theta)+\sin\alpha(\sin(2\phi-2\theta)-i\sin2\theta)\right].
\end{align}
Using these polarizations, the full four-photon Stark shift $\delta\omega^{(4)}(\alpha,\beta,\theta,\phi)$ may then be determined by substituting into Eqs. \ref{eq: Rabis} and \ref{eq:e4all}.
\section{Results}

In our experiments, we employ the four-photon Stark shift effect as a primary means of characterizing the Raman laser beams used for driving quantum gate operations. While earlier works have used the four-photon Shift to perform local single-qubit gate operations \cite{lee2016engineering}, here we use it as a tool for diagnosing a variety of important beam parameters. Below, we calibrate the performance of the four-photon Stark shift in our system before analyzing and optimizing the Raman beam polarizations, spot sizes, and alignments.

\subsection{Beam Power and Polarization}
\begin{figure}[h]
\begin{centering}
\includegraphics[width=14.0 cm,  trim=145 170 70 180, clip]{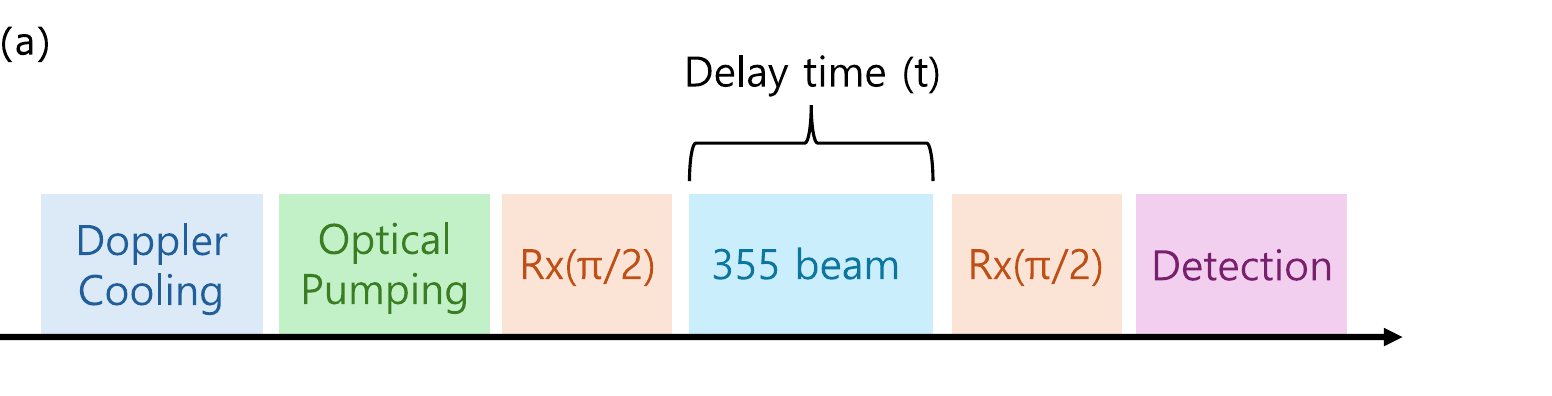}
\includegraphics[width=7.0 cm]{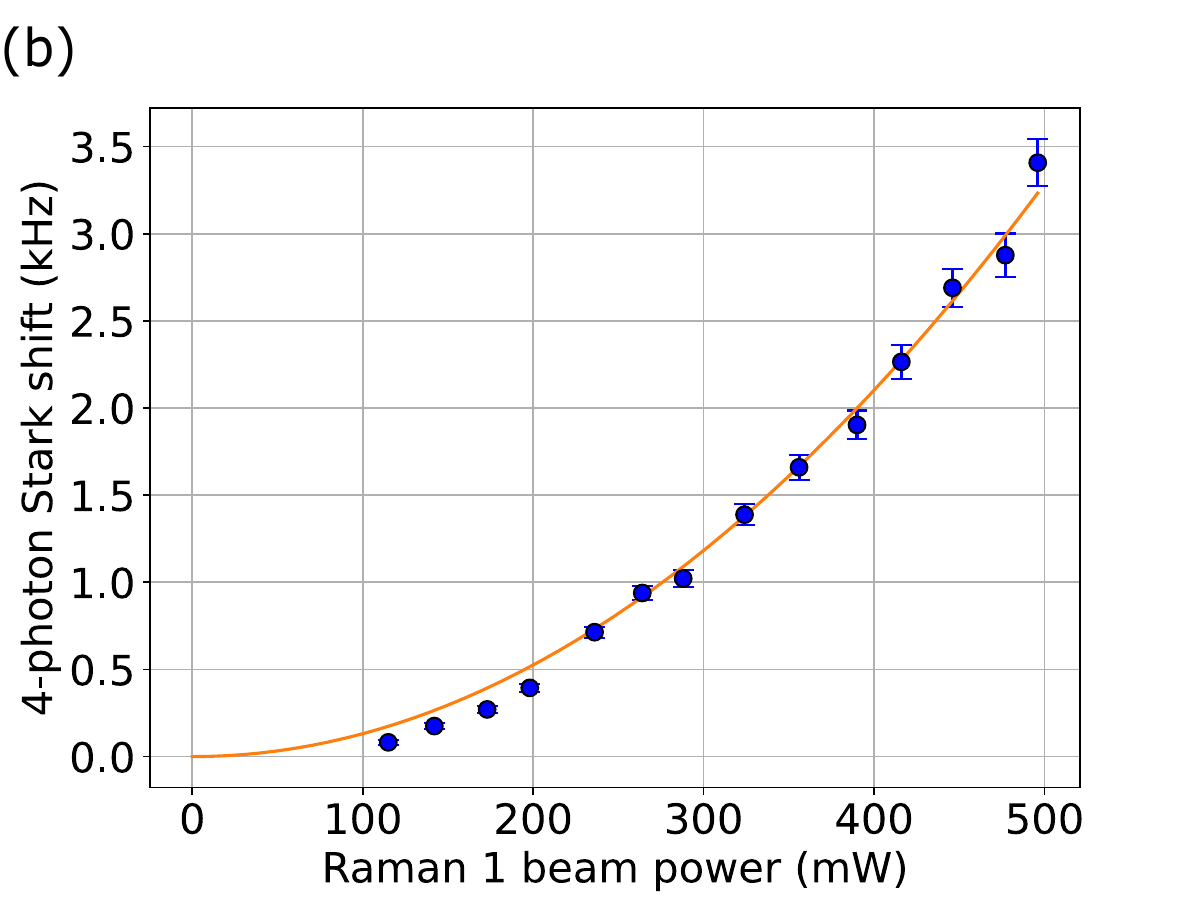}
\includegraphics[width=7.0 cm]{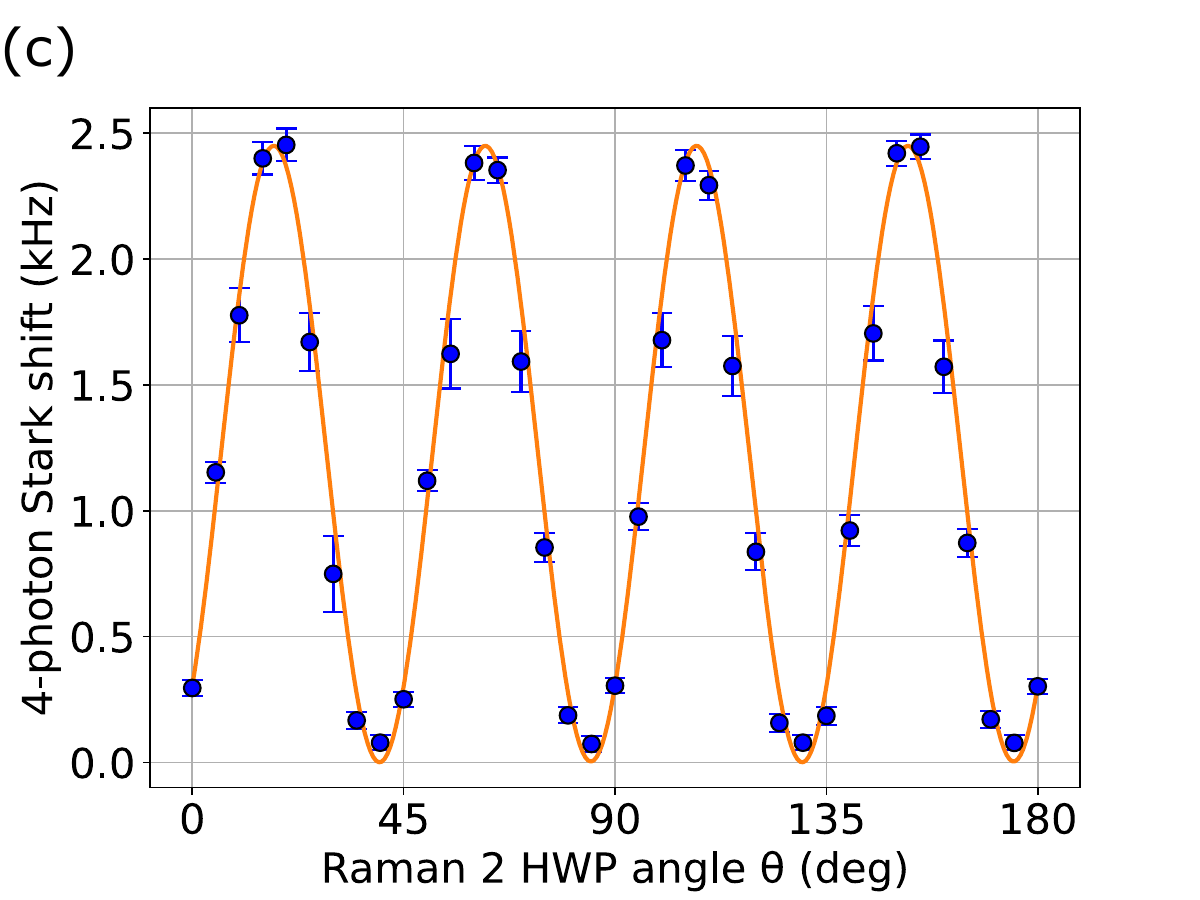}
\end{centering}
\caption{(a) Experimental sequence used to measure the four-photon Stark shift. (b) The four-photon Stark shift exhibits a quadratic dependence (fitted orange line) on the input Raman beam power. (c) The shift also shows an oscillatory dependence as the HWP angle is scanned. The oscillation amplitude, offset, and shape reveal information about the incoming beam polarization and external magnetic field direction. The solid line is a fit to Equation \ref{eq:dw4}, using the polarizations in Equation \ref{eq:genepsilon} as inputs.}
\label{powerandpolariztation}
\end{figure}

All four-photon Stark shift measurements are performed following the pulse sequence shown in Figure \ref{powerandpolariztation}(a). After ion cooling and initialization to the $\ket{\downarrow}$ state, the system is prepared in the equatorial plane of the Bloch sphere using a microwave $\pi/2$ pulse at the qubit transition frequency. A single 355 nm Raman beam is applied for a variable delay time to drive the four-photon shift, after which the Ramsey sequence is completed by a second microwave $\pi/2$ pulse. Oscillations in the detected probabilities of $\ket{\downarrow}$ and $\ket{\uparrow}$ then determine the Stark shift frequency $\delta\omega^{(4)}$. Each Ramsey experiment is repeated 100 times with results averaged together. Systematic uncertainties for each data point account for state preparation and measurement (SPAM) errors of approximately 4\%, intensity noise in the laser and microwave fields of approximately 1\%, and ion-position measurement uncertainty of approximately $\pm1~\mu$m. In addition, fluctuations of our beam polarizations and applied magnetic fields are constrained to $< 3\%$.

Since the Stark shift is expected to be zero in the ideal lin $\perp$ lin beam configuration (see Equation \ref{eq:linpolss}), we use a HWP to purposefully rotate the beam polarization away from horizontal for these experiments. For linear polarizations with both horizontal and vertical components, we expect a non-zero Stark shift that depends quadratically on $\Omega_0$, and therefore, on input beam power. In Figure \ref{powerandpolariztation}(b), we verify this dependence by setting the HWP at angle $\theta = 22.5^\circ$ to the horizontal axis and measuring the Stark shift for increasing 355 nm powers. We note that small deviations from the quadratic dependence are visible at low power, due to the residual two-photon ac Stark Shift.

As the HWP angle is rotated, the observed Stark shift contains diagnostic information about the input Raman beam polarization and external magnetic field direction. This dependence may be calculated by solving Equation \ref{eq:dw4}, using the generalized polarization components in Equation \ref{eq:genepsilon} as inputs to Equation \ref{eq: Rabis}. The result is a four-photon Stark shift $\delta\omega^{(4)}(\alpha,\beta,\theta,\phi)$ that can indicate an imbalance of left- and right-circularly polarized light (corresponding to $\phi \neq 0$) or a non-vertical magnetic field.

We study the Stark shift dependence on HWP angle in Figure \ref{powerandpolariztation}(c), in the case of balanced circular polarization ($\phi=0$) and non-vertical magnetic field ($\alpha, \beta \neq 0$). As the HWP angle is scanned, we observe high-contrast oscillations with minimal Stark shift whenever $\sin^2(4\theta) \approx 0$, as predicted by Eq. \ref{eq:linpolss}. However, we also observe a phase offset from pure $\sin^2(4\theta)$ dependence that is attributable to a non-perfectly vertical magnetic field. Fitting the full Stark shift $\delta\omega^{(4)}(\alpha,\beta,\theta,\phi)$ to the data in Figure \ref{powerandpolariztation}(c) reveals that the magnetic field used in these experiments is misaligned from the $\hat{z}$ axis by an angle $\alpha = 10^\circ$.

\subsection{Beam Intensity Profiles and Alignments}
The speed and stability of single- and two-qubit gates depends directly on the profile and alignment of each Raman beam. In the ideal case, the maximum gate speed is attained when the peak intensity of each Raman beam overlaps with the ion position. In addition, this configuration reduces gate infidelities due to laser pointing instability since the intensity profile of a Gaussian beam varies most slowly near its peak.

Using the four-photon Stark shift, we characterize the intensity profile of each Raman beam and use these measurements to improve their alignments to the ions. As shown in Equation \ref{eq:linpolss} and Figure \ref{powerandpolariztation}(b), we expect the induced four-photon Stark shift to depend on the square of the local Raman beam intensity. To measure this profile, we translate the ion along the linear trap axis and record the Stark shift at each position for each beam. 

\begin{figure}[h]
\begin{centering}
\includegraphics[width=7.0 cm]{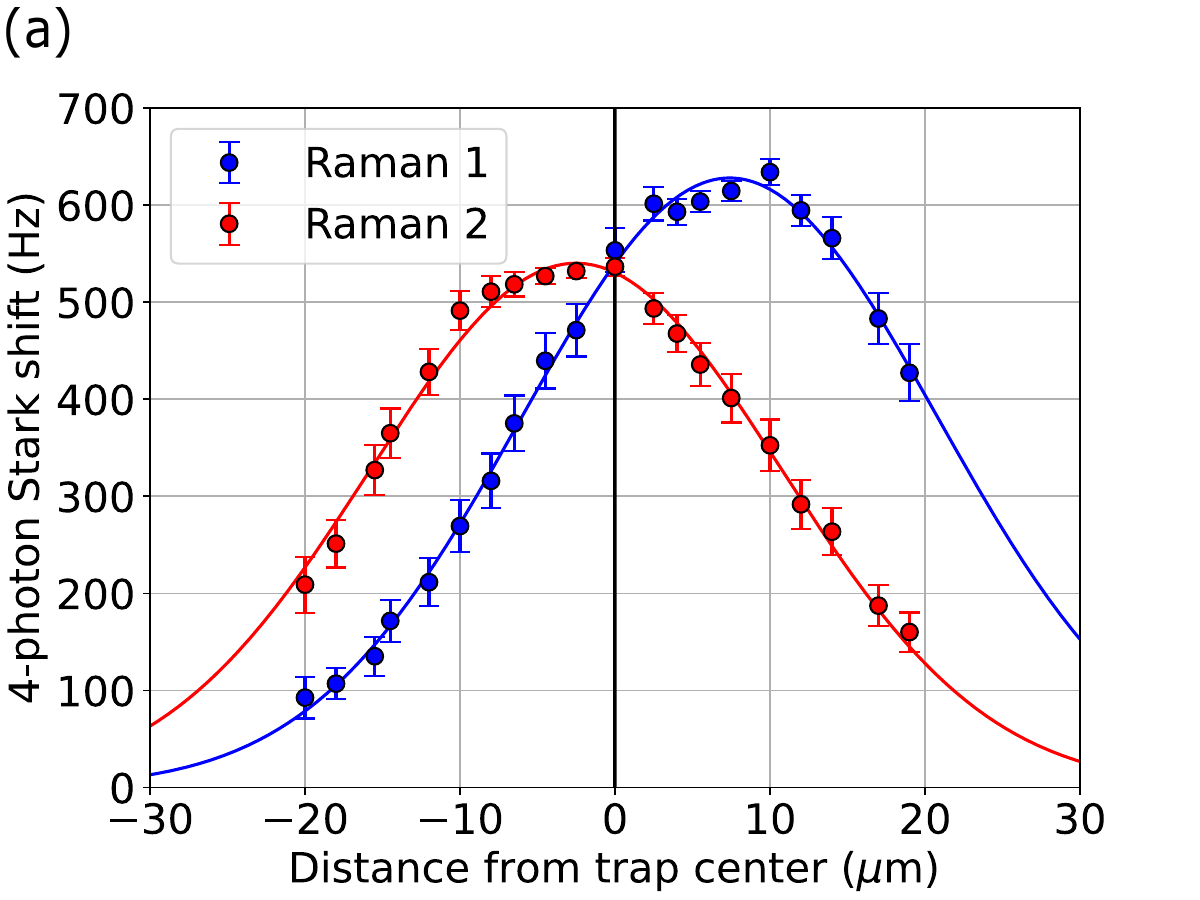}
\includegraphics[width=7.0 cm]{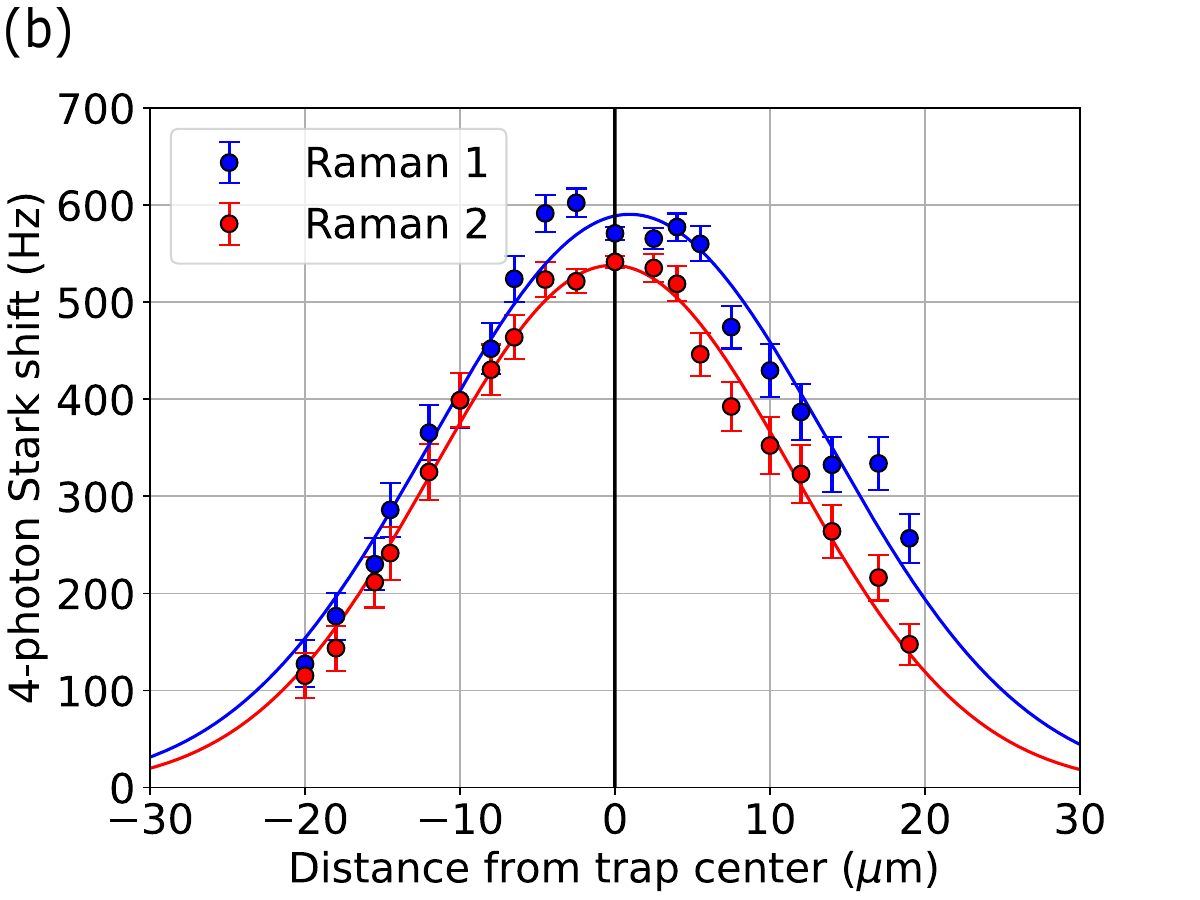}
\end{centering}
\caption{Raman beam laser profiles as measured via the four-photon Stark shift. Panel (a) shows the profiles before alignment optimization; panel (b) is after optimization.}
\label{profile}
\end{figure}   

The four-photon Stark shift for each Raman beam as a function of ion position is shown in Figure \ref{profile}(a). For both beams, we fit a squared Gaussian profile to the data to extract a $1/e^2$ waist. Accounting for the $45^\circ$ projection of each Raman beam along the trap axis, we measure waists of $w_0=27\pm2~\mu$m, which is near our planned beam waist target of $30~\mu$m. Further, the high fit quality confirms that there are no significant beam aberrations or non-Gaussian modes at the focal plane of the ion. However, beam profiles in Figure \ref{profile}(a) highlight that the individual Raman lasers were misaligned to the ion position by as much as 8~$\mu$m. This is corrected in Figure \ref{profile}(b), where both beams are brought into alignment with the ion to sub-$\mu$m accuracy. 

\begin{figure}[h]
\begin{centering}
\hspace*{.7cm}
\includegraphics[width=14 cm, trim=130 100 0 160, clip]{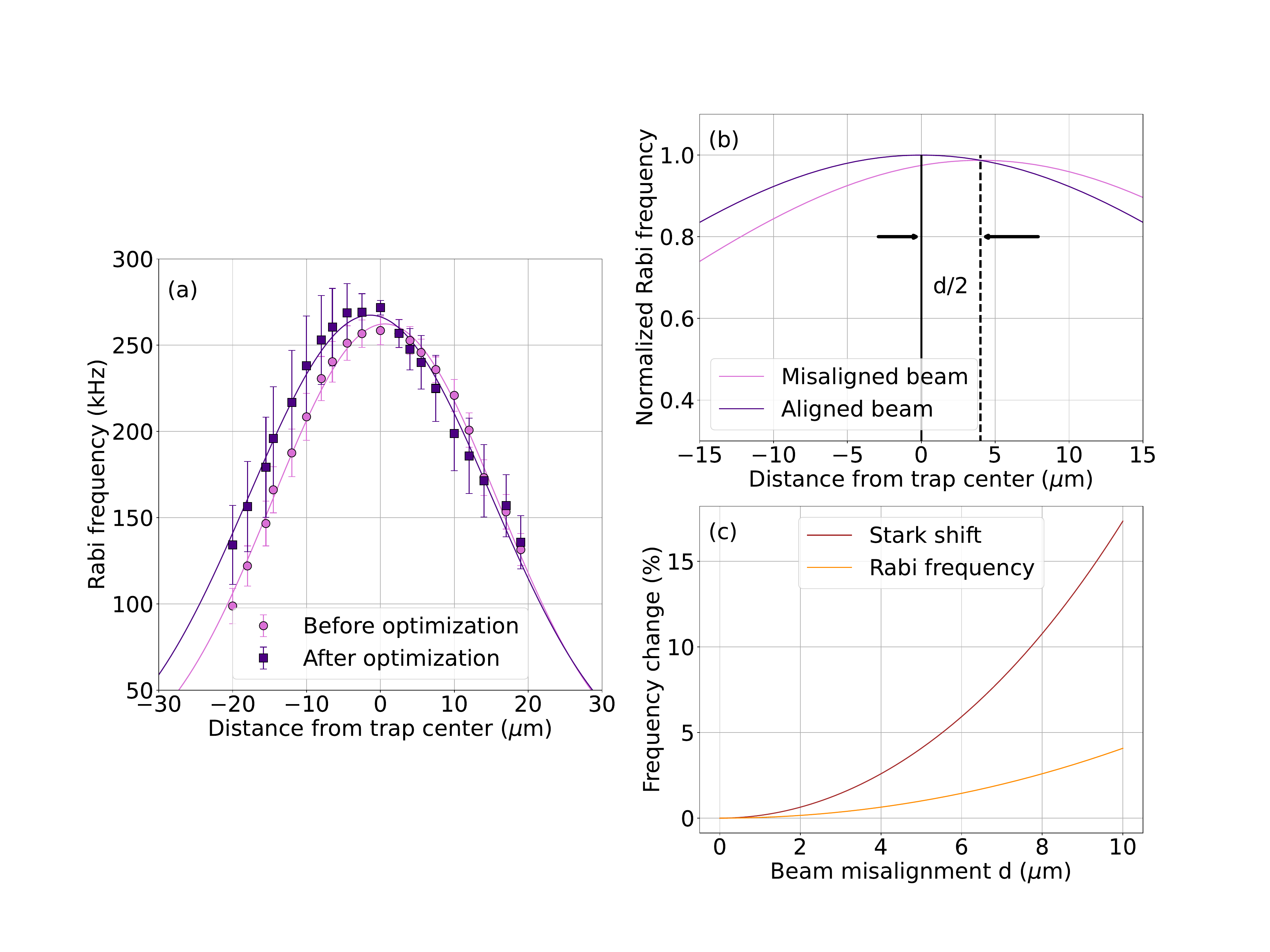}
\end{centering}
\caption{(a) Rabi frequency versus ion position, before and after the alignment optimizations shown in Figure \ref{profile}. Optimization only slightly increases the measured two-photon Rabi frequency. (b) When one Raman beam is misaligned by a distance $d$, the resulting Rabi frequency curve shifts by $d/2$ with little change to its peak amplitude. (c) The four-photon Stark shift signal is much more sensitive to misalignments than the Rabi frequency signal, with increasing sensitivity for larger deviations.}
\label{rabi}
\end{figure}

Although Rabi-frequency mapping or two-photon ac Stark Shift spectroscopy are traditional methods to align Raman beams, we will now show that the four-photon Stark shift measurements provide better sensitivity to alignment deviations. Ensuring high-accuracy alignment is particularly important for reducing pointing instability effects from the Raman beams, which manifest themselves as gate over- or under-rotation errors. In Figure \ref{rabi}(a) we measure the two-photon Rabi frequency versus ion position, before and after optimizing the beam alignments. Although optimization increased the peak Rabi frequency from 260 kHz to 270 kHz, this difference is small and can be challenging to discern given common levels of experimental noise.

The poor sensitivity of Rabi frequencies to beam misalignments can be predicted theoretically. Figure \ref{rabi}(b) shows the calculated Rabi frequencies along the trap axis when the Raman beams are aligned, and also when one Raman beam is misaligned by $d=8~\mu$m. Because the two-photon Rabi frequency $\Omega \sim \sqrt{I_1 I_2}$ is proportional to the geometric mean of Raman beam intensities, the misaligned Rabi peak shifts by $d/2$ (assuming $I_1 = I_2$) and exhibits only a small decrease in amplitude. In contrast, the four-photon Stark shift signal scales as $\delta\omega^{(4)}\sim I^2$ for each beam individually and provides a more sensitive probe of beam misalignments. 

The relative sensitivities of the Stark-shift and Rabi frequency measurements are compared in Figure \ref{rabi}(c). For both approaches, we calculate the fractional change in frequency when one of the Raman beams is misaligned by a distance $d$. In all cases, Stark shift measurements provide a larger change in signal, with a $\sim4\times$ improved sensitivity for our initial misalignment value of $d=8~\mu$m. This theoretical analysis, along with our experimental observations, highlight the advantages of using four-photon Stark shifts to optimize beam alignment.


\section{Discussion and Conclusions}

In this work, we demonstrated the four-photon Stark shift as a technique for characterizing the laser beams used to drive quantum gate operations. After introducing the origins of the effect, we extended its theoretical description to include experimental realities such as imperfect beam polarizations and magnetic field directions. Using the ion as a probe of its environment, we applied the four-photon Stark shift technique to measure Raman laser beam polarizations, spot sizes, and alignments. We further showed how alignment optimizations can be improved using four-photon Stark shift measurements compared to traditional Rabi frequency optimization. Our work demonstrates that ions are not just passive targets for quantum operations, but can actively serve as local sensors to calibrate and improve experiments.

\begin{acknowledgments}
This research was funded by the National Science Foundation under award OSI-2435255 andthe Gordon and Betty Moore Foundation, grant DOI 10.37807/GBMF12963.
\end{acknowledgments}

\bibliography{main}{}

\appendix
\section[\appendixname~\thesection]{}
\label{appxa}
In order to calculate the four-photon Stark shift allowing for arbitrary magnetic fields directions and arbitrary components of linear and circular polarization, we begin by defining an external magnetic field, parameterized by both polar and azimuthal angles $\alpha$ and $\beta$, with respect to the coordinate axes as

\begin{equation}
\label{arbB}
    \vec{B} = B_{o}\left[\sin\alpha\cos\beta\hat{x}+\sin\alpha\sin\beta\hat{y}+\cos\alpha\hat{z}\right]
\end{equation}

\noindent We also assume an input Raman beam that starts with horizontal polarization and is made to pass through QWP at angle $\phi$ and a HWP at angle $\theta$. The QWP and HWP plates are represented as Jones transformations, where the fast axis is with respect to the horizontal direction

\begin{equation}
    QWP(\phi)=e^{-i\pi/4}\begin{pmatrix}
        \cos^2\phi+i\sin^2\phi && (1-i)\sin\phi\cos\phi \\
        (1-i)\sin\phi\cos\phi && \sin^2\phi+i\cos^2\phi
    \end{pmatrix}
\end{equation}

\begin{equation}
    HWP(\theta)=e^{-i\pi/2}
    \begin{pmatrix}
        \cos^2\theta - \sin^2\theta && 2\cos\theta\sin\theta \\
        2 \cos\theta\sin\theta && \sin^2\theta-\cos^2\theta
    \end{pmatrix}
\end{equation}


Because the magnetic field sets the quantization axis of the ion, we need to rotate our polarization vector to be in the ion's frame

\begin{equation} \hat{\epsilon}'=\epsilon_{x'}\hat{x}'+\epsilon_{y'}\hat{y}'+\epsilon_{z'}\hat{z}'
\end{equation}

\noindent Where we take the $z$-direction of our rotated coordinate system ($\hat{z}'$) to be along the magnetic field

\begin{align}
    \nonumber
    \hat{x}'&=\cos\alpha\cos\beta\hat{x}+\cos\alpha\sin\beta\hat{y}-\sin\alpha\hat{z} \\[1em]
    \nonumber
    \hat{y}'&= -\sin\beta\hat{x}+\cos\beta\hat{y} \\[1em]
    \hat{z}'&=\sin\alpha\cos\beta\hat{x}+\sin\alpha\sin\beta\hat{y}+\cos\alpha\hat{z}
\end{align}

\noindent It should be noted that the unit vectors $\hat{x}'$ and $\hat{y}'$ are not unique, as long as the transformed unit vectors form an orthonormal basis. 

We can find the coefficients of the polarization in the ion's frame by taking the inner product of the rotated coordinate system's unit vector with the Raman beam's polarization after the QWP and HWP:

\begin{align}
    \label{eq:dotproducts}
    \nonumber
    \epsilon_{x'} &= \hat{x}'\cdot\hat{\epsilon} = -\frac{1}{\sqrt{2}}\left [\cos\alpha\sin\beta(\cos(2\phi-2\theta)+i\cos2\theta)+\sin\alpha(\sin(2\phi-2\theta)-i\sin2\theta) \right]\\[1em]
    \nonumber
    \epsilon_{y'} &= \hat{y}'\cdot\hat{\epsilon} = -\frac{1}{\sqrt{2}}\cos\beta(\cos(2\phi-2\theta)+i\cos2\theta)\\[1em]
    \epsilon_{z'} &= \hat{z}'\cdot\hat{\epsilon} = -\frac{1}{\sqrt{2}}\left[\sin\alpha\sin\beta(\cos(2\phi-2\theta)+i\cos2\theta)-\cos\alpha(\sin(2\phi-2\theta)-i\sin2\theta) \right]
\end{align}
To match the polarization basis required in Equation \ref{eeq:polbasis} of the main text, we can transform the primed coordinates as



\begin{equation}
    \hat{\epsilon}_{\pm}=\frac{1}{\sqrt{2}}(\hat{x}'\pm i \hat{y}'), \quad \hat{\epsilon}_{\pi} = \hat{z}'
\end{equation}

\noindent We can find the coefficients of the polarization in the $\hat{\epsilon}_{-}$,$\hat{\epsilon}_{\pi}$,$\hat{\epsilon}_{+}$ basis by applying the same process as in \ref{eq:dotproducts} 

\begin{align}
    \nonumber
        \epsilon_- &= -\frac{1}{2}\left[(\cos\alpha\sin\beta-i\cos\beta)(\cos(2\phi-2\theta)+i\cos2\theta)+\sin\alpha(\sin(2\phi-2\theta)-i\sin2\theta)\right] \\[1em]
    \nonumber
    \epsilon_\pi &= -\frac{1}{\sqrt{2}}\left[\sin\alpha\sin\beta(\cos(2\phi-2\theta)+i\cos2\theta)-\cos\alpha(\sin(2\phi-2\theta)-i\sin2\theta)\right]\\[1em]
    \epsilon_+ &= -\frac{1}{2}\left[(\cos\alpha\sin\beta+i\cos\beta)(\cos(2\phi-2\theta)+i\cos2\theta)+\sin\alpha(\sin(2\phi-2\theta)-i\sin2\theta)\right].
\end{align}





\end{document}